\documentclass[prd,superscriptaddress,amsfonts,amssymb,amsmath,showpacs,twocolumn]{revtex4-2}
\usepackage{bm}
\usepackage{amsfonts}
\usepackage{latexsym}
\usepackage[latin1]{inputenc}
\usepackage{graphicx}
\usepackage{amsmath}
\usepackage{palatino}
\usepackage{mathpazo}
\usepackage{textcomp}
\usepackage{float}
\usepackage{booktabs}
\usepackage{dcolumn}
\usepackage{booktabs}
\usepackage{multirow}
\usepackage{hyperref}
\hypersetup{colorlinks,citecolor=blue}
\usepackage{amsmath}
\usepackage{xcolor}
\usepackage{orcidlink}
\usepackage[caption=false]{subfig}
\usepackage{commath}
\def\jnl@style{\it}
\def\aaref@jnl#1{{\jnl@style#1}}

\def\aaref@jnl#1{{\jnl@style#1}}

\def\aj{\aaref@jnl{AJ}}                   
\def\apj{\aaref@jnl{ApJ}}                 
\def\apjl{\aaref@jnl{ApJ}}                
\def\apjs{\aaref@jnl{ApJS}}               
\def\apss{\aaref@jnl{Ap\&SS}}             
\def\aap{\aaref@jnl{A\&A}}                
\def\aapr{\aaref@jnl{A\&A~Rev.}}          
\def\aaps{\aaref@jnl{A\&AS}}              
\def\mnras{\aaref@jnl{Mon.~Not.~Roy.~Astron.~Soc.}}             
\def\prd{\aaref@jnl{Phys.~Rev.~D}}        
\def\prc{\aaref@jnl{Phys.~Rev.~C}}  
\def\prl{\aaref@jnl{Phys.~Rev.~Lett.}}    
\def\qjras{\aaref@jnl{QJRAS}}             
\def\skytel{\aaref@jnl{S\&T}}             
\def\ssr{\aaref@jnl{Space~Sci.~Rev.}}     
\def\zap{\aaref@jnl{ZAp}}                 
\def\nat{\aaref@jnl{Nature}}              
\def\aplett{\aaref@jnl{Astrophys.~Lett.}} 
\def\apspr{\aaref@jnl{Astrophys.~Space~Phys.~Res.}} 
\def\physrep{\aaref@jnl{Phys.~Rep.}}      
\def\physscr{\aaref@jnl{Phys.~Scr}}       
\def\commat{\aaref@jnl{Comm.~Math.~Phys.}}              
\def\science{\aaref@jnl{Science}}               
\def\cqg{\aaref@jnl{Classical Quant.~Grav.}}            
\def\jpcs{\aaref@jnl{JPCS}}                                     
\def\ijmpd{\aaref@jnl{Int.~J.~Mod.~Phys.~D}}                    
\def\grg{\aaref@jnl{Gen.~Relat.~Gravit.}}               
\def\rpp{\aaref@jnl{Rep.~Prog.~Phys.}}          
\def\npa{\aaref@jnl{Nucl.~Phys.~A}}        
\def\lrr{\aaref@jnl{Living Rev.~Rel.}}                   
\def\jcap{\aaref@jnl{J.~Cosmology Astropart.~Phys.}}    
\def\rmp{\aaref@jnl{Rev.~Mod.~Phys.}}   
\def\epjc{\aaref@jnl{Eur.~Phys.~J.~C}}

\allowdisplaybreaks[1]

\begin{document}

\color{black}       

\title{Bouncing cosmologies and stability analysis in symmetric teleparallel $f(Q)$ gravity}

\author{M. Koussour\orcidlink{0000-0002-4188-0572}}
\email[Email: ]{pr.mouhssine@gmail.com}
\affiliation{Department of Physics, University of Hassan II Casablanca, Morocco.}

\author{N. Myrzakulov\orcidlink{0000-0001-8691-9939}}
\email[Email: ]{nmyrzakulov@gmail.com}
\affiliation{L. N. Gumilyov Eurasian National University, Astana 010008,
Kazakhstan.}


\begin{abstract}
This paper is devoted to examining cosmological bouncing scenarios in the framework of the recently proposed symmetric teleparallel gravity (or $f(Q)$ gravity), where the non-metricity scalar $Q$ represents the gravitational interaction. We assume an $f(Q)$ model in the form of $f(Q)=\alpha Q^n$, where $\alpha$ and $n$ are free model parameters. To obtain a bouncing universe, we consider a special form of the scale factor $a(t)$ in terms of cosmic time, specifically $a(t) = (1+\lambda t^2)^{1/3}$, where $\lambda$ is an arbitrary constant. We derive the field equations for the flat FLRW universe and obtain the corresponding exact solution. We investigate the physical behavior of various cosmological parameters such as the deceleration parameter, pressure, and equation of state (EoS) parameter with the energy conditions for our bounce cosmological model. Furthermore, we investigate the behavior of the perturbation terms $\delta_m(t)$ and $\delta(t)$ with respect to cosmic time $t$ using the scalar perturbation approach. We found that although the model exhibits unstable behavior at the beginning for a brief period, it shows mostly stable behavior for most of the time. Finally, we conclude that the EoS parameter crosses the quintom line $\omega=-1$ in the vicinity of the bouncing point $t=0$, which confirms the success of our bounce cosmological model.
\end{abstract}
\maketitle

\tableofcontents
\section{Introduction}

\label{sec1}

General relativity (GR) is a geometric theory of gravity based on Riemannian geometry, which extends Euclid's flat geometry to describe curved surfaces. Together with quantum physics, GR stands as a remarkable achievement in modern physics. This influential theory is highly successful and currently represents our best understanding of gravity. These successes are due to numerous tests and predictions, such as the perihelion advance of Mercury, the deflection of light by the Sun, and the detection of the
gravitational waves, etc \cite{GR1, GR2}. On the other hand, recent
observations in cosmology such as Type Ia Supernovae (SNIa) \cite{SN1, SN2},
Cosmic Microwave Background (CMB) \cite{CMB1, CMB2}, Wilkinson Microwave
Anisotropy Probe (WMAP) data \cite{WMAP1, WMAP2, WMAP9}, Large Scale
Structure (LSS) \cite{LS1, LS2}, and Baryonic Acoustic Oscillations (BAO) 
\cite{BAO1, BAO2} have provided conclusive evidence that our universe has
now entered a phase of accelerated expansion. In addition, the same data supports the conclusion that $95\%$ of the total content of the universe in the form of
two exotic components of energy and matter called Dark Energy (DE) and Dark
Matter (DM), respectively, with only $5\%$ represents ordinary matter in the
form of baryonic matter.

Although GR has provided explanations for many phenomena within the solar
system, these recent observations have thrown this theory into great
trouble. In fact, GR cannot explain many gravitational phenomena on a large
scale in the universe. Thus, GR may not be the definitive theory of gravity, because it is not able to account for the present acceleration of the
universe (or DE), DM, the initial singularity, and the singularity of the
black hole. To interpret the results of recent cosmological data, several
alternatives have recently been proposed. An approach called modified
theories of gravity (MTG), where it is suggested that Einstein's theory of
gravity is invalid on the grand scale of the universe, and the
Einstein-Hilbert action, which describes GR, must be modified to a more
general action. In GR, gravitational interactions are described by Ricci
curvature $R$, as a generalization, it is to replace the Ricci curvature $R$ by an arbitrary function $f\left( R\right) $ and the result is the so-called 
$f\left( R\right) $ gravity \cite{fR}. A second alternative to extending the
Einstein-Hilbert action is to presume a non-minimal coupling between
geometry and matter such as $f\left( R,T\right) $ and $f\left(
R,L_{m}\right) $, where $T$ and $L_{m}$ are the trace of the energy-momentum
tensor and the matter Lagrangian density, respectively \cite{fRL, fRT}.

Since GR is a geometric theory, another approach has been taken, which is to
generalize Riemannian geometry such as Weyl geometry. In Riemannian
geometry, the curvature of space-time is measured by the variation of the
direction of a vector in the parallel transport process, while in Weyl
geometry, the variation of the length of a vector is also taken into
account. This leads to the covariant derivative of the metric tensor $g_{\mu
\nu }$ is non-zero in Weyl geometry, and this is called the non-metricity
i.e. $Q_{\gamma \mu \nu }=\nabla _{\gamma }g_{\mu \nu }$ \cite{Xu}. Another
extension of Weyl geometry is known as Weyl-Cartan geometry where torsion $T$
is introduced. According to this presentation, gravitational interactions
can be described by three concepts: (i) curvature (GR) in which the torsion
and the non-metricity are zero, (ii) torsion (teleparallel gravity) in which
the curvature and the non-metricity are zero, and (iii) non-metricity
(symmetric teleparallel gravity) in which the curvature and the torsion are
zero. As mentioned, the $f\left( R\right) $ gravity is a generalization of
GR, similarly $f\left( T\right) $ gravity is a generalization of
teleparallel gravity \cite{fT}\ and $f\left( Q\right) $ gravity is a
generalization of symmetric teleparallel gravity. In this work, we will
discuss the newly suggested $f(Q)$ gravity in which the non-metricity scalar
describes the gravitational interactions \cite{Jimenez1}. Harko et al.
investigated the coupling matter in the modified $Q$ theory of gravity \cite%
{Harko1}. The growth index of matter perturbations has been analysed in the
background of $f(Q)$ gravity \cite{Khyllep}. The signatures of $f\left(
Q\right) $\ gravity has been analysed in \cite{Frusciante}.

In the literature, there are three scenarios that describe the cosmic
expansion and predict the ultimate fate of the universe. The first scenario
is that there could be so much matter in the observable universe that
despite the observed expansion of gravity, it would bring everything back to
a big crunch. The second scenario is that galaxies recede from each other
and space-time itself expands all the time. The final scenario is the idea
of the oscillating universe, which describes a model of the universe that
alternates between expanding and contracting phases, with big crunch and big
bang between these phases, and is famous as the big bounce, which we ought
to examine in the context of symmetric teleparallel gravity. The big bounce
theory is an attractive cosmological model that describes the origin of the
universe without the initial singularity found in GR because in this theory
the universe passes from contraction to expansion without collapsing on
itself \cite{Haro, Moriconi}. In addition, bouncing cosmology contradicts
the existence of the initial singularity. Thus, this cosmological model is
considered an effective solution to the problem of singularity in the
standard model of the big bang. Several authors have discussed the idea of a
bouncing universe in various contexts such as $f\left( R\right) $, $f\left(
R,T\right) $, $f\left( G\right) $, $f\left( R,G\right) $, and $f\left(
Q\right) $ gravities \cite{Odintsov, Singh, Barros, Bajardi, Bhattacharjee,
Sahoo, BounQ,fRLm_B,fQT_B}. The null energy condition (NEC) is included in most phenomenological models, which makes it difficult to realize a bouncing cosmological model. The NEC, which is the sum of isotropic pressure and energy density of the universe, must be violated for the Hubble rate to increase and the bounce to occur \cite{BounQ1}. However, violating the NEC can introduce instability issues such as the Belinski-Khalatnikov-Lifshitz (BKL) instability \cite{BounQ2}. This instability occurs when the anisotropic energy density of space-time increases faster than that of the bouncing agent during the contracting phase, resulting in an unstable background evolution. Therefore, the matter bounce scenario suffers from two significant flaws: (i) BKL instability; and (ii) in the perturbation evolution, a large tensor-to-scalar ratio implies that the scalar and tensor perturbations have similar amplitudes. An exact matter bounce scenario with a single scalar field results in an essentially scale-invariant power spectrum \cite{BounQ3}.

The paper is ordered as follows: In Sec.~\ref{sec2}, we present an overview
of $f(Q)$ gravity theory in the framework of a flat FLRW universe. In Sec.~\ref{sec3}, we briefly discuss the energy conditions in $f(Q)$ gravity. Next, we
consider some cosmological solutions to obtain the bouncing universe in Sec.~\ref{sec4}. The behavior of some cosmological parameters of the bouncing $f(Q)$ model, such as the pressure, and EoS parameter, are discussed in Sec.~\ref{sec5}. In Sec.~\ref{sec6}, we show the evolution of the stability analysis of the model. Finally, we conclude with our results in Sec.~\ref{sec7}.

\section{$f(Q)$ gravity theory}
\label{sec2}

As it is well known, the metric tensor $g_{\mu \nu }$ in GR is a
generalization of the concept of gravitational potentials in Newton's
theory. Generally, the metric tensor is used to determine distances,
volumes, and angles while the affine connection $\Sigma {^{\gamma }}_{\mu
\nu }$ is used as a basic tool in the parallel transport process and
covariant derivatives. In the differential geometry of the Weyl-Cartan type
with the presence of torsion $T$ and non-metricity $Q$\ terms, the most
general affine connection can be given in terms of all possible
contributions as \cite{Xu} 
\begin{equation}
\Sigma {^{\gamma }}_{\mu \nu }={\Gamma ^{\gamma }}_{\mu \nu }+K{^{\gamma }}%
_{\mu \nu }+{L^{\gamma }}_{\mu \nu },  \label{WC}
\end{equation}%
where ${\Gamma ^{\gamma }}_{\mu \nu }$, $K{^{\gamma }}_{\mu \nu }$ and ${%
L^{\gamma }}_{\mu \nu }$ are the Levi--Civita connection, the contorsion
tensor and the disformation tensor, respectively. 
\begin{equation}
{\Gamma ^{\gamma }}_{\mu \nu }\equiv \frac{1}{2}g^{\gamma \sigma }\left(
\partial _{\mu }g_{\sigma \nu }+\partial _{\nu }g_{\sigma \mu }-\partial
_{\sigma }g_{\mu \nu }\right) ,
\label{LC}
\end{equation}%
\begin{equation}
K{^{\gamma }}_{\mu \nu }\equiv \frac{1}{2}g^{\gamma \sigma }\left( T_{\mu
\sigma \nu }+T_{\nu \sigma \mu }+T_{\sigma \mu \nu }\right) ,
\end{equation}%
\begin{equation}
{L^{\gamma }}_{\mu \nu }\equiv \frac{1}{2}g^{\gamma \sigma }\left( Q_{\nu
\mu \sigma }+Q_{\mu \nu \sigma }-Q_{\gamma \mu \nu }\right) .
\end{equation}

The torsion tensor is determined by the antisymmetric part of $\Sigma {%
^{\gamma }}_{\mu \nu }$, while the non-metricity tensor from the covariant
derivative of the metric $g_{\mu \nu }$ as, 
\begin{equation}
{T^{\gamma }}_{\mu \nu }\equiv 2\Sigma {^{\gamma }}_{[\mu \nu ]}\text{, \ \ }%
Q_{\gamma \mu \nu }=-\nabla _{\gamma }g_{\mu \nu }\neq 0,
\end{equation}%
and $Q_{\gamma \mu \nu }$ in terms of the most general connection is given
as, 
\begin{equation}
Q_{\gamma \mu \nu }=-\partial _{\gamma }g_{\mu \nu }+g_{\nu \sigma }\Sigma {%
^{\sigma }}_{\mu \gamma }+g_{\sigma \mu }\Sigma {^{\sigma }}_{\nu \gamma }.
\end{equation}

GR in which gravitational interactions are outlined by the concept of
curvature can be obtained from the above description by the absence of both
the contorsion term and the disformation term i.e. $K{^{\gamma }}_{\mu \nu
}= $ ${L^{\gamma }}_{\mu \nu }=0$. In addition, depending on the form of the
connection, two different other theories that are equivalent to GR can be
constructed, namely: TEGR (Teleparallel Equivalent to General Relativity)
i.e. ${L^{\gamma }}_{\mu \nu }=0$ and
STEGR (Symmetric Teleparallel Equivalent to General Relativity) i.e. ${K^{\gamma }}_{\mu \nu }=0$. In this work, we
have focused on the last presentation of GR i.e. STEGR. If space-time is
considered flat with zero torsion, it must match to a pure coordinate
transformation from the trivial connection as exhibit in \cite{Jimenez1}.
More clearly, the connection can be parameterised as%
\begin{equation}
\Sigma {^{\gamma }}_{\mu \beta }=\frac{\partial x^{\gamma }}{\partial \xi
^{\rho }}\partial _{\mu }\partial _{\beta }\xi ^{\rho },  \label{par}
\end{equation}

It is good to point out in Eq. (\ref{par}) that $\xi ^{\gamma }=$ $\xi
^{\gamma }\left( x^{\mu }\right) $ is an invertible relation and $\frac{%
\partial x^{\gamma }}{\partial \xi ^{\rho }}$ is the inverse of the
corresponding Jacobian \cite{Jimenez2}. Thus, it is always possible to get a
coordinate system in which the general connection is zero i.e. $\Sigma {%
^{\gamma }}_{\mu \nu }=0$ \cite{Jimenez1}. Also, the curvature tensor is
zero which makes the overall geometry of space-time flat as the Weitzenb\"ock
geometry. The previous condition is known as \textit{coincident gauge} and
the covariant\ derivative $\nabla _{\gamma }$ reduces to the partial
derivative $\partial _{\gamma }$. Thus in the coincident gauge coordinate,
can be gained $Q_{\gamma \mu \nu }=-\partial _{\gamma }g_{\mu \nu }$. It is
clear from the above discussion that the Levi-Civita connection ${\Gamma
^{\gamma }}_{\mu \nu }$ can be written in terms of the disformation tensor ${%
L^{\gamma }}_{\mu \nu }$ as ${\Gamma ^{\gamma }}_{\mu \nu }=-{L^{\gamma }}%
_{\mu \nu }$. The action that corresponds to the STEGR is described by
\begin{equation}
S_{STEGR}=\int \sqrt{-g}d^{4}x\left[\frac{1}{2}(-Q)+L_{m}\right] .
\end{equation}

The $f(Q)$ theory of gravity is a generalization of the STEGR in which the
extended action is given by \cite{Jimenez2} 
\begin{equation}
S=\int \sqrt{-g}d^{4}x\left[ \frac{1}{2}f(Q)+L_{m}\right] ,  \label{action}
\end{equation}%
where $f(Q)$ represents an arbitrary function of the non-metricity scalar $Q$%
, $g$ is the determinant of the metric tensor $g_{\mu \nu }$, and $L_{m}$ is
the matter Lagrangian density. In addition, from the above action, GR can be
reproduced for the option of function in the form $f\left( Q\right) =-Q$,
i.e. for this option we recover the known as STEGR \cite{Lazkoz}. Now, owing
to the symmetricity of $g_{\mu \nu }$ there are only two independent traces
procured from the non-metricity term $Q_{\gamma \mu \nu }$ specifically, 
\begin{equation}
Q_{\gamma }={{Q_{\gamma }}^{\mu }}_{\mu }\,,\qquad \widetilde{Q}_{\gamma }={%
Q^{\mu }}_{\gamma \mu }\,.
\end{equation}

In addition, it is useful to introduce the superpotential tensor i.e.
non-metricity conjugate given by 
\begin{equation}
4{P^{\gamma }}_{\mu \nu }=-{Q^{\gamma }}_{\mu \nu }+2Q{_{(\mu \;\;\nu
)}^{\;\;\;\gamma }}+Q^{\gamma }g_{\mu \nu }-\widetilde{Q}^{\gamma }g_{\mu
\nu }-\delta _{\;(\mu }^{\gamma }Q_{\nu )}\,.
\end{equation}

Then the trace of the non-metricity tensor can be obtained as 
\begin{equation}
Q=-Q_{\gamma \mu \nu }P^{\gamma \mu \nu }\,.
\end{equation}

The Riemann curvature tensor is defined as
\begin{equation}\label{2h}
R^\gamma_{\: \beta\mu\nu} = 2\partial_{[\mu} \Sigma^\gamma_{\: \nu]\beta} + 2\Sigma^\gamma_{\: [\mu \mid \lambda \mid}\Sigma^\lambda_{\nu]\beta}.
\end{equation}

Using the affine connection given in Eq. \eqref{WC}, we obtain
\begin{equation}\label{2i}
R^\gamma_{\: \beta\mu\nu} = \mathring{R}^\gamma_{\: \beta\mu\nu} + \mathring{\nabla}_\mu X^\gamma_{\: \nu \beta} - \mathring{\nabla}_\nu X^\gamma_{\: \mu \beta} + X^\gamma_{\: \mu\rho} X^\rho_{\: \nu\beta} - X^\gamma_{\: \nu \rho} X^\rho_{\: \mu\beta}.
\end{equation}

In this context, $\mathring{R}^\gamma_{\: \beta\mu\nu}$ and $\mathring{\nabla}$ are defined with respect to the Levi-Civita connection \eqref{LC}, and $X{^{\gamma }}%
_{\mu \nu }=K{^{\gamma }}%
_{\mu \nu }+{L^{\gamma }}_{\mu \nu }$. By applying appropriate contractions to the curvature term and imposing the torsion-free constraint $T{^{\gamma }}%
_{\mu \nu }=0$ in Eq. \eqref{2i}, we obtain
\begin{equation}\label{2j}
R=\mathring{R}-Q + \mathring{\nabla}_\gamma \left(Q^\gamma-\tilde{Q}^\gamma \right),   
\end{equation}
where $\mathring{R}$ represents the usual Ricci scalar calculated using the Levi-Civita connection. By imposing the teleparallel constraint $R=0$, we achieve curvature-free teleparallel geometries, and consequently, Eq. \eqref{2j} simplifies to
\begin{equation}\label{2k}
\mathring{R}=Q - \mathring{\nabla}_\gamma \left(Q^\gamma-\tilde{Q}^\gamma \right).   
\end{equation}

From Eq. (\ref{2k}), it is evident that the Ricci scalar, when calculated using the Levi-Civita connection, differs from the non-metricity scalar $Q$ by a total derivative. Applying the generalized Stokes' theorem, this total derivative can be converted into a boundary term \cite{ADe}. Consequently, the Lagrangian density changes by a boundary term, indicating that $Q$ is equivalent to $\mathring{R}$. Thus, $Q$ provides a comparable description of GR.

By varying the action in Eq. (\ref{action}) with respect to the metric
tensor $g_{\mu \nu }$, we get the field equations for the $f\left( Q\right) $%
\ symmetric teleparallel gravity as, 
\begin{equation}\label{F}
\begin{split}
\frac{2}{\sqrt{-g}}\nabla_\gamma (\sqrt{-g}f_Q P^\gamma\:_{\mu\nu}) 
&+ \frac{1}{2}g_{\mu\nu}f \\
&+ f_Q(P_{\mu\gamma\beta}Q_\nu\:^{\gamma\beta} - 2Q_{\gamma\beta\mu}P^{\gamma\beta}\:_\nu) = -T_{\mu\nu}.
\end{split}
\end{equation}
where $f_{Q}={df}/{dQ}$ and $\nabla _{\mu }$ denotes the covariant
derivative. While the first two terms are manifestly symmetric, the third term can be shown to be symmetric as well. This ensures that the field equations of $f(Q)$ gravity are symmetric, preserving local Lorentz invariance and confirming that no additional degrees of freedom are introduced when analyzing perturbations \cite{Heisenberg}. Furthermore, the energy-momentum tensor for the perfect fluid
matter of the universe is given by 
\begin{equation}
T_{\mu \nu }=-\frac{2}{\sqrt{-g}}\frac{\delta (\sqrt{-g}\mathcal{L}_{m})}{%
\delta g^{\mu \nu }}.
\end{equation}

In addition, by varying the action with regard to the connexion, we find 
\begin{equation}
\nabla ^{\mu }\nabla ^{\nu }\left( \sqrt{-g}\,f_{Q}\,P^{\gamma }\;_{\mu \nu
}\right) =0.
\end{equation}

Taking into account the cosmological principle which reads that our universe
is homogeneous and isotropic on large scales. In this work, we assume the
following flat Friedmann-Lemaitre-Robertson-Walker (FLRW) metric, 
\begin{equation}
ds^{2}=-dt^{2}+a^{2}(t)\left[ dx^{2}+dy^{2}+dz^{2}\right] ,  \label{FLRW}
\end{equation}%
where $a(t)$ is the scale factor of the universe that measures the cosmic
expansion at a time $t$. The non-metricity scalar corresponding to the flat
FLRW metric is obtained as 
\begin{equation}
Q=6H^{2},
\end{equation}%
where $H$ is the Hubble parameter which measures the rate of expansion of
the universe.

In the case of a universe filled with perfect fluid type matter-content, the
energy-momentum tensor is defined as%
\begin{equation}
T_{\mu \nu }=(p+\rho )u_{\mu }u_{\nu }+pg_{\mu \nu },
\end{equation}%
where $p$ and $\rho $ represent the isotropic pressure and the energy
density of the universe, respectively. Here, $u^{\mu }=\left( 1,0,0,0\right) 
$ are components of the four velocities of the perfect fluid.

Thus, the modified Friedmann equations that describe the dynamics of the
universe in $f\left( Q\right) $\ symmetric teleparallel gravity read as 
\begin{equation}
3H^{2}=\frac{1}{2f_{Q}}\left( -\rho +\frac{f}{2}\right) ,  \label{F1}
\end{equation}%
\begin{equation}
\dot{H}+3H^{2}+\frac{\dot{f}_{Q}}{f_{Q}}H=\frac{1}{2f_{Q}}\left( p+\frac{f}{2%
}\right) ,  \label{F2}
\end{equation}%
where the dot $(\overset{.}{})$ denotes the derivative with regard to the
cosmic time $t$. Especially, for $f(Q)=-Q$ we retrieve the standard GR
Friedmann's equations \cite{Harko1}, as mentioned above, this specific
option for the functional form of the function $f(Q)$ is the STEGR limit of
the theory. The continuity equation of the energy-momentum tensor writes 
\begin{equation}
\dot{\rho}+3H(\rho +p)=0.
\end{equation}

Using Eqs. (\ref{F1}) and (\ref{F2}), we obtain the expressions of the
energy density of the universe $\rho $ and the isotropic pressure $p$,
respectively as 
\begin{equation}
\rho =\frac{f}{2}-6H^{2}f_{Q},  \label{F22}
\end{equation}%
\begin{equation}
p=\left( \dot{H}+3H^{2}+\frac{\dot{f_{Q}}}{f_{Q}}H\right) 2f_{Q}-\frac{f}{2}.
\label{F33}
\end{equation}

Again, by using Eqs. (\ref{F1}) and (\ref{F2}) we can rewrite the
cosmological equations similar to the standard Friedmann equations in GR, by
adding the concept of an effective energy density $\overline{\rho }$ and an
effective isotropic pressure $\overline{p}$ as%
\begin{equation}
3H^{2}=\overline{\rho }\,=-\frac{1}{2f_{Q}}\left( \rho -\frac{f}{2}\right) ,
\label{F222}
\end{equation}%
\begin{equation}
2\dot{H}+3H^{2}=-\overline{p}\,=-\frac{2\dot{f_{Q}}}{f_{Q}}H+\frac{1}{2f_{Q}}%
\left( \rho +2p+\frac{f}{2}\right) .  \label{F333}
\end{equation}

Moreover, the gravitational action (\ref{action}) is reduced to the standard
Hilbert-Einstein form in the limiting case $f\left( Q\right) =-Q$. For this
choice, Eqs. (\ref{F222}) and (\ref{F333}) reduce to the standard Friedmann
equations of GR, $3H^{2}=\rho $, and $2\dot{H}+3H^{2}=-p$, respectively.

\section{Energy conditions}

\label{sec3}

The energy conditions (ECs) are a set of simple constraints on different
linear combinations of the energy density of the universe and isotropic
pressure. These conditions show that the energy density of the universe
cannot be negative and that gravity is always attractive and have many
applications in theoretical cosmology. For example the ECs play an important
role in GR as they help to prove the theorems about the presence of the
singularity of space-time and black holes \cite{Wald}. In the context of
this work, the ECs are used for two reasons: to verify the bouncing cosmic
scenario and to predict the acceleration phase of the universe. The ECs can
be obtained from the Raychaudhury equations, which are given as \cite%
{Raychaudhuri, Nojiri2, Ehlers} 
\begin{equation}
\frac{d\theta }{d\tau }=-\frac{1}{3}\theta ^{2}-\sigma _{\mu \nu }\sigma
^{\mu \nu }+\omega _{\mu \nu }\omega ^{\mu \nu }-R_{\mu \nu }u^{\mu }u^{\nu
}\,,  \label{R1}
\end{equation}%
\begin{equation}
\frac{d\theta }{d\tau }=-\frac{1}{2}\theta ^{2}-\sigma _{\mu \nu }\sigma
^{\mu \nu }+\omega _{\mu \nu }\omega ^{\mu \nu }-R_{\mu \nu }n^{\mu }n^{\nu
}\,,  \label{R2}
\end{equation}%
where $n^{\mu }$, $\theta $, $\omega _{\mu \nu }$ and $\sigma ^{\mu \nu }$
are the null vector, the expansion factor, the rotation and the shear
associated with the vector field $u^{\mu }$, respectively. In Weyl geometry
with the existence of non-metricity scalar $Q$, the Raychaudhury equations
take various forms, for more details see \cite{Arora}. For
attractive gravity, Eqs. (\ref{R1}) and (\ref{R2}) fulfill the following
conditions 
\begin{align}
R_{\mu \nu }u^{\mu }u^{\nu }& \geq 0\,, \\
R_{\mu \nu }n^{\mu }n^{\nu }& \geq 0\,.
\end{align}

Thus, if we examine the perfect fluid distribution of cosmological matter,
the ECs for $f(Q)$ gravity are given as follows,

\begin{itemize}
\item WEC (Weak energy condition): if $\overline{\rho }\geq 0$, $\overline{%
\rho }+\overline{p}\geq 0$.

\item NEC (Null energy condition): if $\overline{\rho }+\overline{p}\geq 0$.

\item DEC (Dominant energy condition): if $\overline{\rho }\geq 0$, $|%
\overline{p}|\leq \overline{\rho }$.

\item SEC (Strong energy condition): if $\overline{\rho }+3\overline{p}\geq
0 $.
\end{itemize}

By taking Eqs. (\ref{F222}) and (\ref{F333}) in the WEC, NEC, and DEC
constraints, we can demonstrate that

\begin{itemize}
\item WEC: if $\rho \geq 0$, $\rho +p\geq 0$.

\item NEC: if $\rho +p\geq 0$.

\item DEC: if $\rho \geq 0$, $|p|\leq \rho $.
\end{itemize}

These findings are consistent with the results obtained by Capozziello et al. \cite%
{Capozziello1}. In the case of the SEC, we obtain
\begin{equation}
\rho +3\,p-6\,\dot{f}_{Q}\,H+f\geq 0\,.
\end{equation}

\section{Bouncing cosmological solutions}

\label{sec4}

In this section, we will discuss one of the cosmological solutions that
produce a bouncing universe. First of all, in order to construct a
successful bouncing dark energy model in standard cosmology, some necessary
conditions are given as follows

\begin{itemize}
\item The first condition is a violation of the null energy condition (NEC)
in the vicinity of the bouncing point, which is equivalent in the standard
FLRW universe $\overset{.}{H}=-4\pi G\rho \left( 1+\omega \right) >0$.

\item The second condition is that, in the phase of contraction of the
universe, the scale factor $a\left( t\right) $ decreases with cosmic time $t$
i.e. $\overset{.}{a}\left( t\right) <0$\ and Hubble parameter $H\left(
t\right) <0$. Whereas, in the phase of expansion of the universe, the scale
factor should increase with cosmic time $t$ i.e. $\overset{.}{a}\left(
t\right) >0$\ and Hubble parameter $H\left( t\right) >0$. Further, $\overset{%
.}{a}\left( t\right) =0$ and Hubble parameter $H\left( t\right) =0$ bouncing
point.

\item Lastly, the equation of the state (EoS) parameter $\omega $\ crosses
the quintom line (phantom divide) $\omega =-1$ in the vicinity of the
bouncing point $t=0$.
\end{itemize}

Taking into account the conditions above, we assume the scale factor of the
form%
\begin{equation}
a\left( t\right) =\left( 1+\lambda t^{2}\right) ^{\frac{1}{3}},
\label{SF}
\end{equation}%
where $\lambda $ is a free model parameter. The choice of the specific form of the scale factor used in the present work is motivated by both theoretical and observational considerations. The scale factor in Eq. (\ref{SF}) is chosen in such a way that it satisfies the following conditions: (i) it is finite and positive for all values of time, (ii) it reaches a non-zero minimum value at $t=0$, corresponding to the bouncing point, and thus provides a description of the origin of the universe without the initial singularity. This is a notable advantage of the present model over other models that rely on the existence of an initial singularity. This form of the scale factor satisfies all the above conditions and ensures the physical viability of the model. The quadratic term in the scale factor is introduced to obtain a bouncing universe, as it allows the universe to undergo a phase of contraction followed by an expansion. In addition, this form of the scale factor has been previously used in various works to study bouncing cosmologies and has been shown to provide physically meaningful results \cite{SF1, SF2}. The corresponding Hubble
parameter $H\left( t\right) $ can be obtained as%
\begin{equation}
H\left( t\right) =\frac{\overset{.}{a}}{a}=\frac{2\lambda t}{3\left(
1+\lambda t^{2}\right) }.  \label{H}
\end{equation}

The deceleration parameter can be obtained by the relation%
\begin{equation}
q\left( t\right) =-1+\frac{d}{dt}\left( \frac{1}{H\left( t\right) }\right) ,
\label{q}
\end{equation}

Using Eqs. (\ref{H}) and (\ref{q}), the deceleration parameter of our model
is derived as%
\begin{equation}
q\left( t\right) =\frac{1}{2}-\frac{3}{2\lambda t^{2}}.
\end{equation}%
\begin{figure}[h]
\centerline{\includegraphics[scale=0.66]{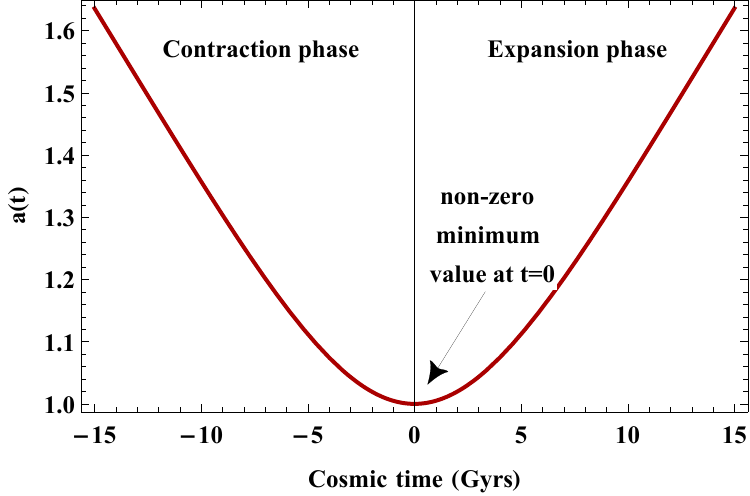}}
\caption{Evolution of the scale factor versus cosmic time with $\protect%
\lambda =0.015$.}
\label{fig_a}
\end{figure}
\begin{figure}[h]
\centerline{\includegraphics[scale=0.66]{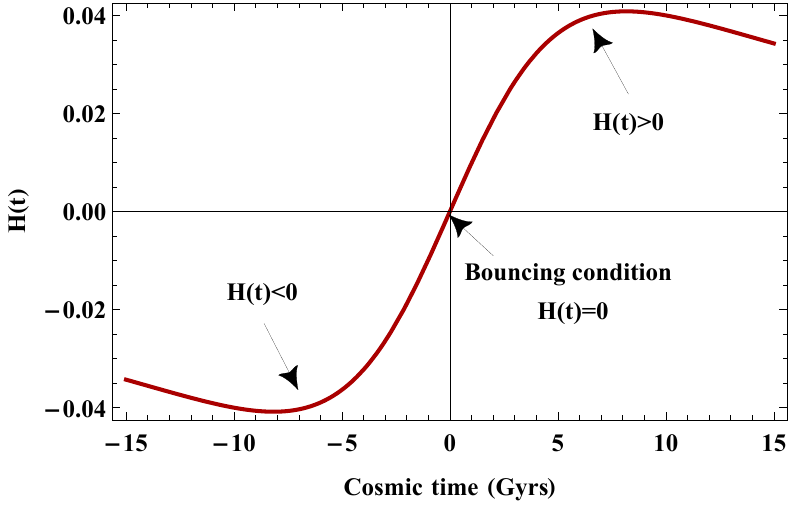}}
\caption{Evolution of the Hubble parameter versus cosmic time with $\protect%
\lambda =0.015$.}
\label{fig_H}
\end{figure}
\begin{figure}[h]
\centerline{\includegraphics[scale=0.65]{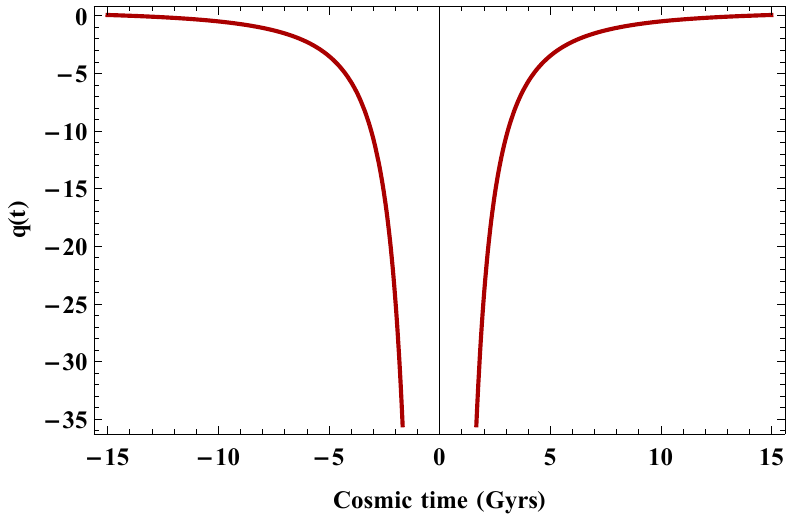}}
\caption{Evolution of the deceleration parameter versus cosmic time with $%
\protect\lambda =0.015$.}
\label{fig_q}
\end{figure}

From Fig. \ref{fig_a}, it is clear that in the contracting universe, the
scale factor is a monotonically decreasing function with cosmic time $t$
i.e. $\overset{.}{a}\left( t\right) <0$, while in the case of the expansion
of the universe, the scale factor is an increasing function with cosmic time 
$t$ i.e. $\overset{.}{a}\left( t\right) >0$. Further, we can see that the
scale factor of the universe reaches to a non-zero minimum value $a\left(
t\right) =1$ at the transition point $t=0$. It is noted that the choice of the scale factor satisfies the required conditions and provides a bouncing cosmology scenario. The specific values of $\lambda$ are chosen to produce physically meaningful results. The spatial volume of the
universe is given as $V\left( t\right) =a^{3}\left( t\right) $. Hence, from
this equation, we can observe that the spatial volume of the universe
decreases before the bounce and starts to increase after the bounce. In the
bouncing universe, the contraction and expansion of the universe can be
described with the help of the Hubble parameter $H\left( t\right) $. Fig. %
\ref{fig_H} indicates two phases of the Hubble parameter $H\left( t\right)
<0 $ for $t<0$\ (contraction) and $H\left( t\right) >0$ for $t>0$\
(expansion) with the bouncing condition satisfied i.e. at $t=0$, $H\left(
t\right) =0$. Hence, we observe that our cosmological model contracts before
the bounce and begin to expand after the bounce.

The deceleration parameter $q\left( t\right) $ is another important tool to
explain the dynamics of the universe. The positive values of the
deceleration parameter ($q>0$) exhibit the deceleration phase of the
universe, while the negative values of $q<0$ point out the acceleration
phase of the universe. From Fig.\ref{fig_q}, it is clear that the
deceleration parameter has a symmetrical behavior at the bouncing point $t=0$%
. Further it is important to check that the deceleration parameter has a
negative value for both the expanding and contracting universes. Finally,
this negative behavior may be consistent with recent observational data
showing that the expansion of the current universe has entered an
accelerating phase. 

\section{Cosmological $f\left( Q\right) $ model}

\label{sec5}

In this section, we discuss the assumed bouncing solutions via a
cosmological model in $f\left( Q\right) $ symmetric teleparallel gravity. In
addition, For a detailed interpretation of the proposed bouncing model, we
need to investigate other required conditions such as energy density,
pressure EoS parameter and null energy condition. For our investigation of
the bouncing cosmological model, we consider the following $f(Q)$ functional
form%
\begin{equation}
f\left( Q\right) =\alpha Q^{n},
\end{equation}%
where $\alpha \neq 0$ and $n$ are free parameters of the model. The choice of the power-law form of the $f(Q)$ function has been used in previous studies and is motivated by its simplicity \cite{Koussour1,Koussour2}. The specific values of $n$ and $\alpha$ are chosen to satisfy the physical constraints and produce consistent cosmological scenarios.

Now by using Eqs. (\ref{F22}) and (\ref{F33}) for the proposed cosmological
model and with the help of the bouncing solutions, we obtained the following
expressions for the energy density of the universe and the isotropic
pressure,%
\begin{widetext}
\begin{equation}
\rho \left( t\right) =\alpha \left( -2^{3n-1}\right) 3^{-n}(2n-1)\left( 
\frac{3H\left( t\right) }{2}\right) ^{2n},
\end{equation}%
and%

\begin{equation}
p\left( t\right) =-\frac{1}{\lambda t^{2}}\left[ \alpha
2^{3n-1}3^{-n}(2n-1)\left( \frac{3H\left( t\right) }{2}\right) ^{2n}\left(
n\left( \lambda t^{2}-1\right) -\lambda t^{2}\right) \right] ,
\end{equation}%
\end{widetext}
respectively. In addition, the EoS parameter plays a critical role in describing the bouncing universe.
For our analysis, the EoS parameter $\omega \left( t\right) $ can be
obtained as%
\begin{equation}
\omega \left( t\right) =\frac{p\left( t\right) }{\rho \left( t\right) }=-%
\frac{n}{\lambda t^{2}}+n-1.
\end{equation}

The energy conditions for our specific form of $f(Q)$ gravity are
\begin{widetext}
\begin{equation}
\text{NEC}\Longleftrightarrow \rho +p=-\frac{1}{\lambda t^{2}}\left[ \alpha
2^{3n-1}3^{-n}n(2n-1)\left( \lambda t^{2}-1\right) \left( \frac{3H\left(
t\right) }{2}\right) ^{2n}\right] \geq 0,
\end{equation}%
\begin{equation}
\text{DEC}\Longleftrightarrow \rho -p=\frac{1}{\lambda t^{2}}\left[ \alpha
2^{3n-1}3^{-n}(2n-1)\left( \frac{3H\left( t\right) }{2}\right) ^{2n}\left(
n\left( \lambda t^{2}-1\right) -2\lambda t^{2}\right) \right] \geq 0,
\end{equation}%
\begin{equation}
\text{SEC}\Longleftrightarrow \rho +3\,p-6\,\dot{f}_{Q}\,H+f=\frac{1}{%
\lambda t^{2}}\left[ \alpha 2^{3n-1}3^{-n}n\left( \lambda t^{2}+3\right)
\left( \frac{3H\left( t\right) }{2}\right) ^{2n}\right] \geq 0.
\end{equation}
\end{widetext}

\begin{figure}[h]
\centerline{\includegraphics[scale=0.65]{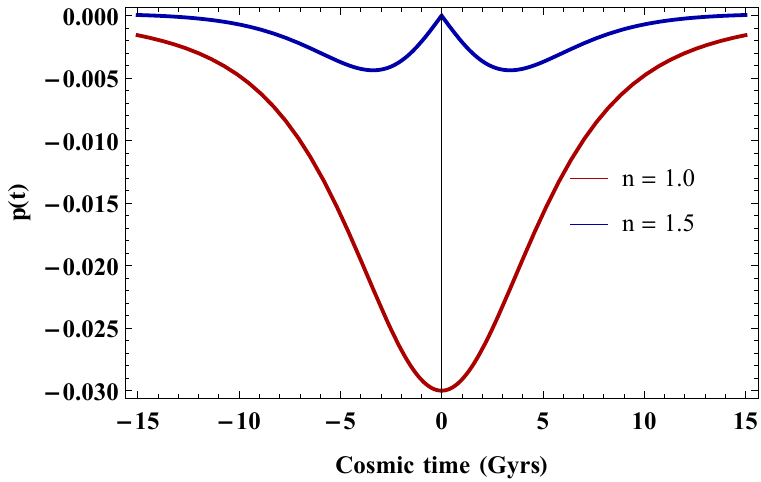}}
\caption{Evolution of the pressure versus cosmic time with $\protect\alpha %
=-1.5$.}
\label{fig_p}
\end{figure}

\begin{figure}[h]
\centerline{\includegraphics[scale=0.65]{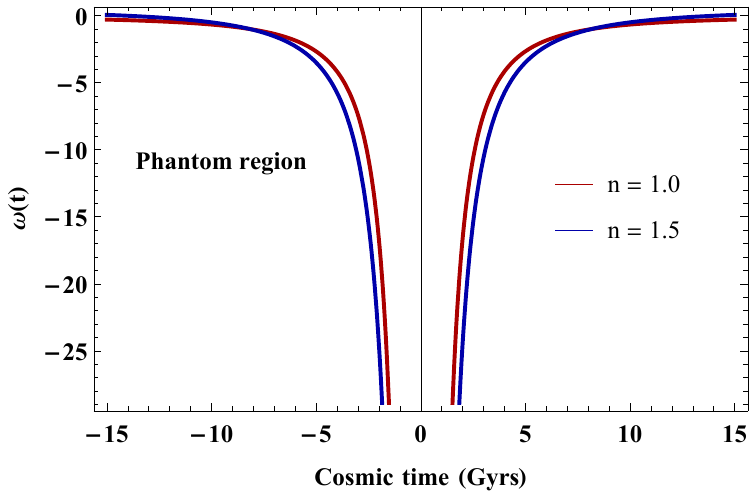}}
\caption{Evolution of the EoS parameter versus cosmic time with $\protect%
\alpha =-1.5$.}
\label{fig_EoS}
\end{figure}

\begin{figure}[h]
\centerline{\includegraphics[scale=0.65]{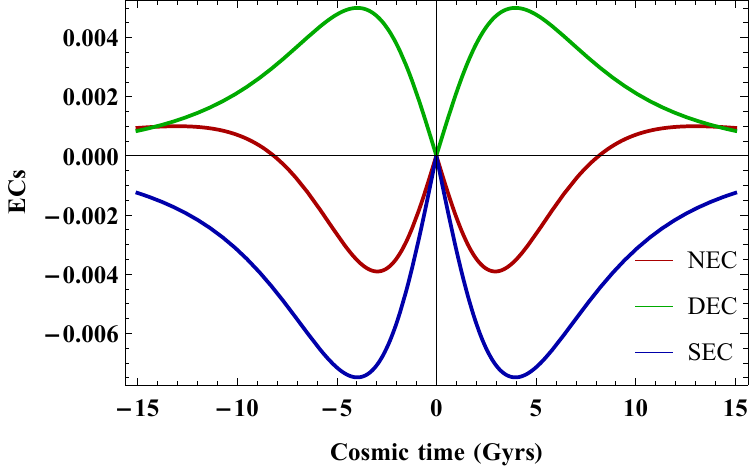}}
\caption{Evolution of the energy conditions versus cosmic time with $\protect%
\alpha =-1.5$ ($n=1.5$).}
\label{fig_ECs2}
\end{figure}

Fig. \ref{fig_p}
indicates that the bouncing solutions exhibit the negative isotropic
pressure for the all range of cosmic time before and after the bouncing
point, Thus, the negative isotropic pressure makes the cosmological bouncing
scenarios a candidate for cosmic acceleration. From Fig. \ref{fig_EoS} it is clear that the EoS parameter for both cases $%
n=1$ and $n=1.5$ crosses the phantom divide i.e. $\omega <-1$ in the
vicinity of the bouncing point $t=0$, which is a very strong criterion for a
successful bouncing cosmological model. Moreover, there is another condition
for the bouncing universe model to be successful, which is that the criteria
for violation of NEC must be satisfied near the bouncing point $t=0$. To
check this, the behavior of all energy conditions is described in Figs. \ref%
{fig_ECs1} and \ref{fig_ECs2} for both cases $n=1$ and $n=1.5$,
respectively. From these figures, we can see that both NEC and SEC are
violated in the vicinity of the bouncing point $t=0$ while the DEC is
fulfilled. The violation of NEC is an important criterion for obtaining the
bouncing universe as mentioned above, while the violation of SEC is a
requirement to obtain the acceleration of the universe due to the presence
of exotic matter. Lastly, it can be said that our cosmological model
satisfies all the fundamental criteria for the bouncing universe in
symmetric teleparallel gravity and also predicts the scenario of the
accelerating universe.

\begin{figure}[h]
\centerline{\includegraphics[scale=0.65]{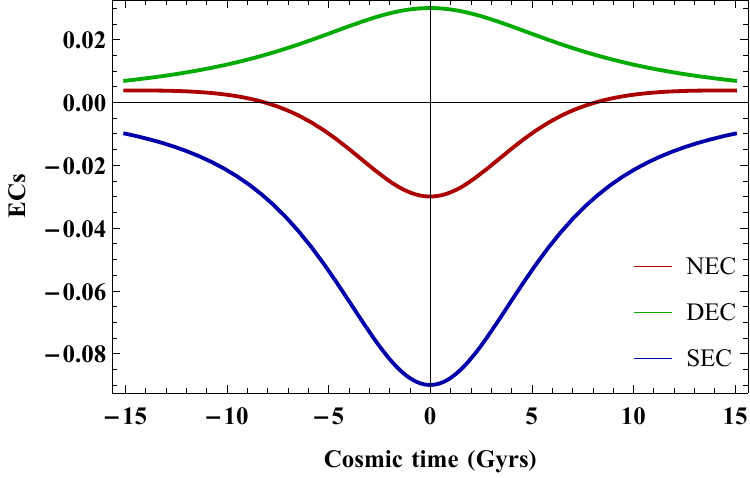}}
\caption{Evolution of the energy conditions versus cosmic time with $\protect%
\alpha =-1.5$ ($n=1$).}
\label{fig_ECs1}
\end{figure}

\section{Linear scalar perturbations}
\label{sec6}
In this section, we will discuss the stability of our cosmological model under homogeneous linear perturbations. Specifically, we define the first-order perturbation for both the Hubble parameter and the density parameter as \cite{Farrugia/2016,Dombriz/2012,Anagnost/2021},
\begin{align} 
\widetilde{H}(t) & = H(t)(1+\delta)\\
\widetilde{\rho}(t) & = \rho(t)(1+\delta_{m}).  
\end{align}

The perturbed Hubble and density parameter are represented by $\widetilde{H}(t)$ and $\widetilde{\rho}(t)$, respectively, while $\delta$ and $\delta_{m}$ correspond to the perturbation terms. In addition, we can express the perturbed $f$ and $f_{Q}$ as $\delta f= f_{Q} \delta Q$ and $\delta f_{Q}= f_{QQ} \delta Q$ with $\delta Q= 12 H \delta H$. Substituting these expressions into the continuity equation and Eq. \eqref{F22}, we obtain the following equations:
\begin{align}
Q \left(f_{Q}+2 Q f_{QQ}\right)\delta &= -\rho \delta_{m},\\
\dot{\delta_{m}} + 3 H (1+\omega) \delta & =0.
\end{align}

By solving the aforementioned equations for $\delta$ and $\delta_{m}$, we obtain
\begin{equation}
\dot{\delta_{m}}- \frac{3 H(1+\omega)\rho}{Q(f_{Q}+2 Qf_{QQ})} \delta_{m}=0.
\end{equation}

After using Eq. \eqref{F33} to simplify the previous equation, the solution can be expressed as follows:
\begin{align}
\delta_{m} & = \delta_{m_{0}} H\\
\delta & = \delta_{0}\frac{\dot{H}}{H}.
\end{align}

The constant $\delta_{m_{0}}$ is introduced and we set $\delta_{0}= -\frac{\delta_{m_{0}}}{3(1+\omega)}$. The solution is obtained for our cosmological model, which is given by:
\begin{align}
\delta_{m}(t) & = \frac{2 \delta_{m_{0}} \lambda  t}{3 \lambda  t^2+3}\\
\delta(t) & = \frac{\delta_{m_{0}} \lambda  t}{3 \lambda  n t^2+3 n}.
\end{align}

\begin{figure}[h]
\centerline{\includegraphics[scale=0.65]{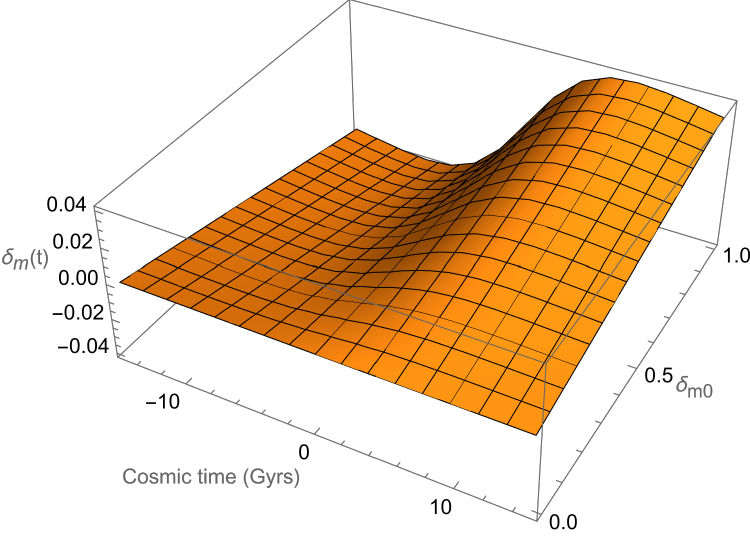}}
\caption{Evolution of the perturbation term $\delta_{m}(t)$ versus cosmic time.}
\label{F_deltam}
\end{figure}

\begin{figure}[h]
\centerline{\includegraphics[scale=0.65]{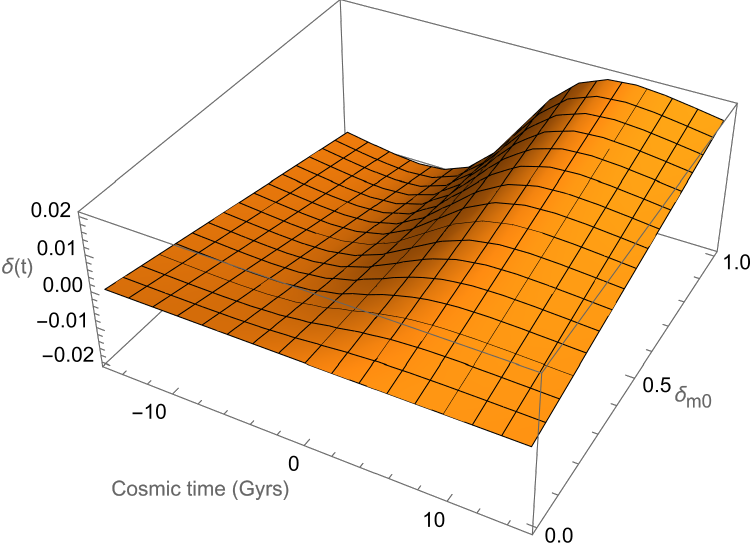}}
\caption{Evolution of the perturbation term $\delta(t)$ versus cosmic time $(n=1)$.}
\label{F_delta1}
\end{figure}

\begin{figure}[h]
\centerline{\includegraphics[scale=0.65]{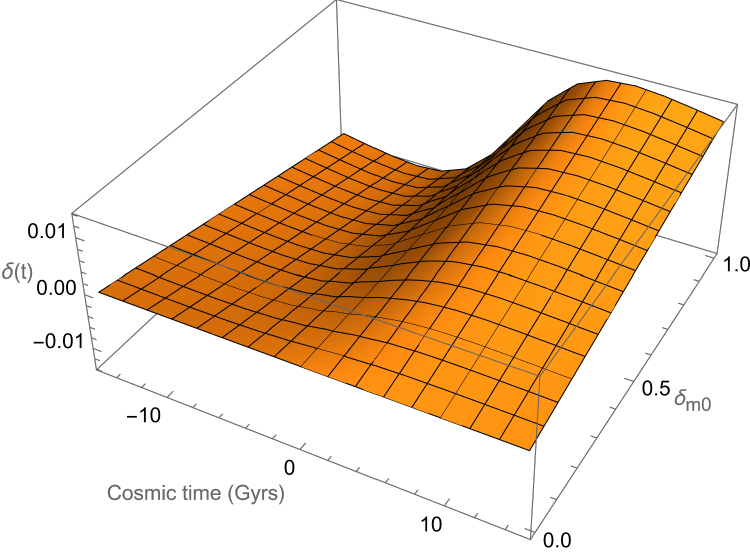}}
\caption{Evolution of the perturbation term $\delta(t)$ versus cosmic time $(n=1.5)$.}
\label{F_delta2}
\end{figure}

The evolution of the perturbation terms $\delta_{m}(t)$ and $\delta(t)$ as a function of cosmic time $t$ is illustrated in Figs. \ref{F_deltam}, \ref{F_delta1}, and \ref{F_delta2} for our cosmological models. We can see that the behavior of both perturbation terms is similar for both values of the parameter $n$. At early times, both $\delta(t)$ and $\delta_m(t)$ increase before reaching a maximum and then decreasing towards zero. This behavior indicates the growth of perturbations during the contracting phase and their subsequent decay during the expanding phase. After the bouncing point, both perturbation terms approach zero, indicating that the perturbations have been stabilized and the universe has returned to a homogeneous and isotropic state. Furthermore, the stability analysis shows that the bouncing cosmology model considered in this paper is stable under scalar perturbations, which is a desirable property for a viable cosmological model.

\section{Conclusions}

\label{sec7}

Bouncing cosmology offers a promising alternative to address the singularity problem and limitations of the inflationary paradigm. It provides a non-singular beginning for the universe and makes distinct predictions that can be tested through observations such as CMB anisotropies \cite{Agullo}, LSS formation \cite{Li}, and primordial gravitational waves \cite{Bergeron}. Future detectors and surveys will be crucial for validation. Bouncing cosmologies can also produce non-Gaussianities and unique behaviors near the bounce point \cite{Agullo/2021}, like the EoS parameter crossing the quintom line. In this work, we investigated the bouncing behavior of the universe in the
framework of $f\left( Q\right) $ symmetric teleparallel gravity theory in
which the non-metricity scalar $Q$ represents the gravitational interaction.
We considered a $f\left( Q\right) $ model in the form of $f\left( Q\right)
=\alpha Q^{n}$, where $\alpha $ and $n$ are free model parameters. Next, to
obtain the corresponding exact solution of the field equations in the FLRW
universe, we proposed a special form of scale factor $a\left( t\right) $ in
terms of cosmic time, specifically, $a\left( t\right) =\left( 1+\lambda
t^{2}\right) ^{\frac{1}{3}}$, where $\lambda $ is an arbitrary constant. In
addition, we have investigated the physical behavior of various cosmological
parameters which shows the bouncing scenario in our cosmological model.

Fig.\ref{fig_a} indicates that the scale factor for our cosmological is a
monotonically decreasing function with cosmic time $t$ i.e. $\overset{.}{a}%
\left( t\right) <0$ in the contracting universe, while in the case of the
expansion of the universe, it is an increasing function with cosmic time $t$
i.e. $\overset{.}{a}\left( t\right) >0$. Further, we have obtained the scale
factor of the universe reaches to a non-zero minimum value $a\left( t\right)
=1$ at the transition point $t=0$. The evolution profile of the Hubble
parameter in Fig. \ref{fig_H} indicates two phases of our model i.e. $%
H\left( t\right) <0$ for $t<0$\ (contraction) and $H\left( t\right) >0$ for $%
t>0$\ (expansion) with the bouncing condition satisfied i.e. at $t=0$, $%
H\left( t\right) =0$. From the figure, we observed that our cosmological
model contracts before the bounce and begin to expand after the bounce.
Furthermore, the deceleration parameter is presented in Fig.\ref{fig_q}
indicates the symmetrical behavior at the bouncing point $t=0$. It is
important to noted that the deceleration parameter has a negative value for
both the expanding and contracting universes.

We have verified the behavior of the EoS parameter, which represents the
phantom behavior ($\omega <-1$) of our model (see Fig. \ref{fig_EoS}) for
both cases $n=1$ and $n=1.5$, which leads to the success of our bounce
cosmological model. Moreover, from Figs. \ref{fig_ECs1} and \ref{fig_ECs2}
we found that NEC and SEC are violated in the vicinity of the bouncing point 
$t=0$ for both cases $n=1$ and $n=1.5$. The violation of NEC satisfies the
bouncing criteria while the violation of SEC depicts the existence of exotic
matter in the universe. Furthermore, we investigated the behavior of the perturbation terms $\delta_m(t)$ and $\delta(t)$ with respect to cosmic time $t$ using the scalar perturbation approach. We have obtained that although the model exhibits unstable behavior at the beginning for a brief period, it shows mostly stable behavior for most of the time. Therefore, we conclude that our presented cosmological
model is a bouncing model in symmetric teleparallel gravity which is
coherent with the models examined by several authors \cite{Odintsov, Singh,
Barros, Bajardi, Bhattacharjee, Sahoo,BounQ}. In both the present work and the work by Mandal et al. \cite{BounQ}, cosmological bouncing scenarios in the framework of symmetric teleparallel gravity have been examined. However, the specific form of the $f(Q)$ function and scale factor used in the two studies differ. The present work employs a power-law form of the $f(Q)$ function and a scale factor with a different form than that used by Mandal. In addition, the analysis of cosmological parameters and adherence to energy conditions yield different results due to these differences. Therefore, although the two studies share some similarities, they offer distinct perspectives on cosmological bouncing scenarios in symmetric teleparallel gravity.

It is important to note that the matter content of the universe is a complex and ongoing area of research. While our proposed model may provide insights into the behavior of the universe in a bouncing scenario, it is a theoretical construct that may not necessarily reflect the physical reality of the universe. We have assumed a specific form of $f(Q)$ and a scale factor to investigate the bouncing scenario, and while this may fix the nature of the matter present in the universe, it is important to acknowledge the limitations of our model. Further investigations are needed to explore the physical realism of the matter content in bouncing cosmological models, and we encourage researchers to explore alternative forms of $f(Q)$ and scale factors to investigate the behavior of the universe in a bouncing scenario.

\textbf{Acknowledgments}
This research is funded by the Science Committee of the Ministry of Science and Higher Education of the Republic of Kazakhstan (Grant No. AP23483654).

\textbf{Data availability} There are no new data associated with this article


\end{document}